\documentclass[conference]{IEEEtran}
\usepackage{amsmath,amssymb,times}
\usepackage{multirow,epsfig,cite,psfrag}
\usepackage{times, amsmath}
\usepackage{stfloats}
\usepackage[table,xcdraw]{xcolor}
\usepackage{dsfont}
\usepackage{multicol, blindtext}
\usepackage{breqn}
\usepackage{color}
\usepackage{enumerate}
\usepackage{amsfonts}
\usepackage{amsthm}
\usepackage{amsmath}
\usepackage{array}
\usepackage{bbm}
\usepackage{cite}
\usepackage{dsfont}
\usepackage{epsfig}
\usepackage{float}
\usepackage{algorithm}
\usepackage[T1]{fontenc}
\usepackage{graphicx}
\usepackage{indentfirst}
\usepackage{multicol}
\usepackage{subfigure}
\usepackage{url}
\usepackage{multirow}
\usepackage{algpseudocode}
\usepackage{color}
\usepackage{epstopdf}
\usepackage{algpseudocode}
\usepackage[normalem]{ulem}
\usepackage{float}
\usepackage[nolist]{acronym}
\usepackage{booktabs}
\usepackage{subfigure}
\usepackage{graphicx}
\usepackage{flushend}

\usepackage{graphicx}
\usepackage{caption}
\usepackage{algorithm}
\usepackage{algpseudocode}
\begin{document}
\title{Fast Uplink Grant for Machine Type Communications: Challenges and Opportunities}
	\author{\IEEEauthorblockN{Samad Ali\IEEEauthorrefmark{1}, Nandana Rajatheva\IEEEauthorrefmark{1}, and Walid Saad\IEEEauthorrefmark{2} }
		\IEEEauthorblockA{\IEEEauthorrefmark{1}
			Center for Wireless Communications (CWC), University of Oulu, Oulu, Finland,\\
			Emails: samad.ali@oulu.fi, nandana.rajatheva@oulu.fi.}
		\IEEEauthorblockA{\IEEEauthorrefmark{2}
			Wireless@VT, Bradley Department of Electrical and Computer Engineering, Virginia Tech, Blacksburg, VA, USA,\\
			Email: walids@vt.edu.}
\vspace{-1cm}
		\thanks{This research was supported by the U.S. National Science Foundation under Grants OAC-1541105 and OAC-1541069.\vspace{-0.5cm}}
	}
\maketitle
\vspace{-2cm}
\begin{abstract}
The notion of a fast uplink grant is emerging as a promising solution for enabling massive machine type communications (MTCs) in the Internet of Things over cellular networks. By using the fast uplink grant, machine type devices (MTD) will no longer require random access (RA) channels to send scheduling requests. Instead, uplink resources can be actively allocated to MTDs by the base station. In this paper, the challenges and opportunities for adopting the fast uplink grant to support MTCs are investigated. First, the fundamentals of fast uplink grant and its advantages over conventional access schemes: i) fully scheduled with RA process and ii) uncoordinated access, are presented. Then, the key challenges that include the prediction of set of MTDs with data to transmit, as well as the optimal scheduling of MTDs, are exposed. To overcome these challenges, a two-stage approach that includes traffic prediction and optimized scheduling is proposed. For this approach, various solutions for source traffic prediction for periodic MTD traffic are reviewed and novel methods for event-driven traffic prediction are proposed. For optimal allocation of uplink grants, new solutions based on advanced machine learning methods are presented. By using the proposed solutions, the fast uplink grant has the potential to enable cellular networks to support massive MTCs and effectively reduce the signaling overhead and overcome the delay and congestion challenges of conventional RA schemes.
\end{abstract}

\section{Introduction} \label{sec:introduction}
\begin{figure}
	\centering
	\includegraphics[width=8cm]{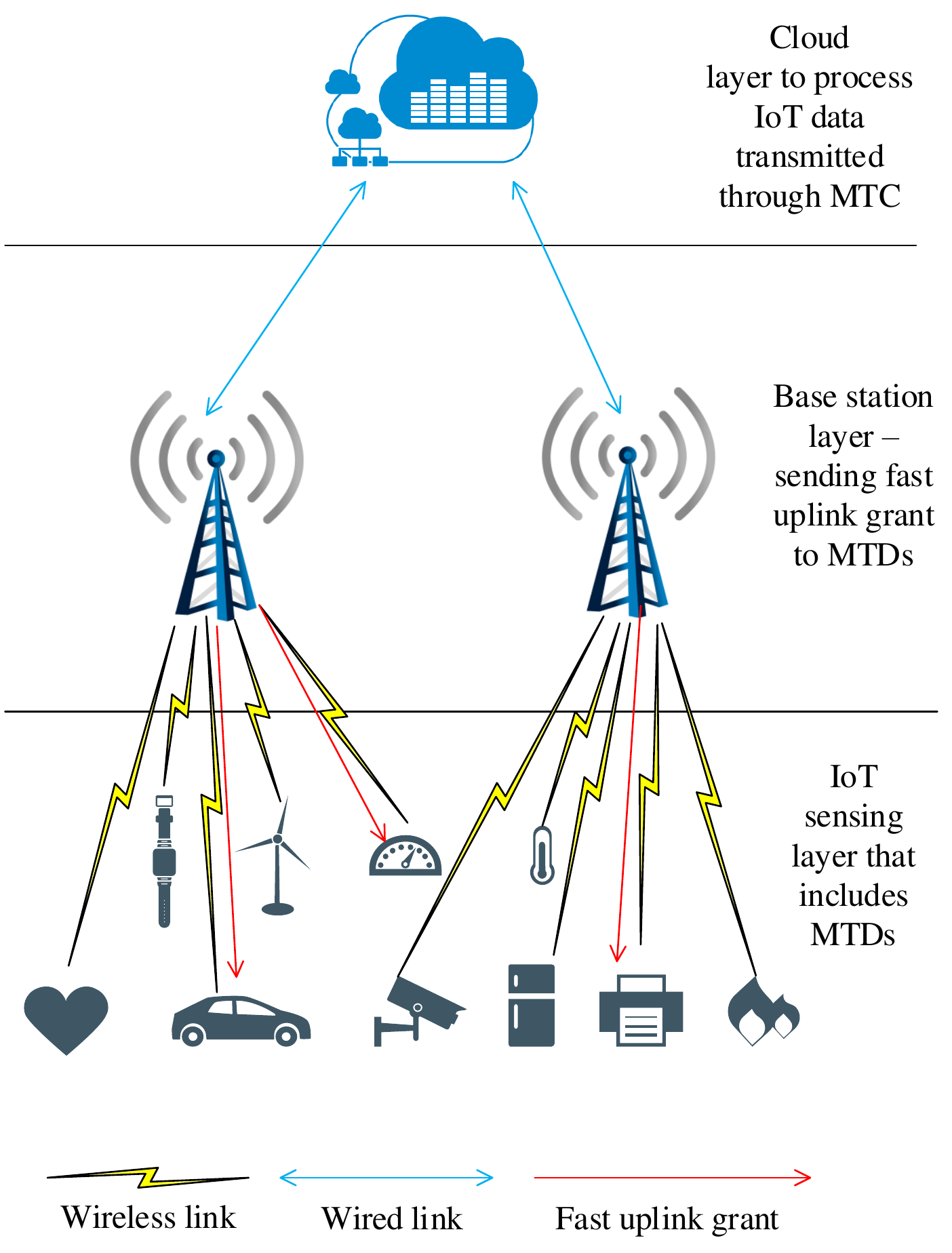}
	\caption{An illustration of an  example IoT environment in which multiple MTDs communicate with BSs connected to a cloud-based gateway.}\label{IoTGeneral}
\end{figure}

Realizing the smart cities vision hinges on the introduction of effective wireless solutions that can provide pervasive connectivity across an Internet of Things (IoT) environment~\cite{IoTforSmartCity} that integrates both human type devices, such as smartphones, and machine type devices (MTDs), such as drones, sensors, vehicles, and actuators. While cellular networks provide an appealing solution for IoT connectivity, existing networks have been designed with a primary focus on providing high data rates to a small number of human type devices, in the downlink. However, as shown in Fig. \ref{IoTGeneral}, IoT applications will rely on a massive number of MTDs that generate small data packets~\cite{dawy2017toward} that are mostly transmitted in the uplink direction, towards a central controller, such as a base station (BS). Such massive machine type communications (MTCs) in the IoT will lead to a major paradigm shift for cellular networks. For instance, beyond its uplink-centered nature, MTC in the IoT will also differ from conventional human type communications by the heterogeneous quality-of-service (QoS) requirements of of the its IoT applications, in terms of latency and reliability, two metrics that are seen as key enablers for IoT applications such as smart grids, autonomous vehicles, factory automation, and e-health. Clearly, supporting such uplink-centric MTCs, with heterogeneous QoS needs will pose major challenges for cellular networks that range from QoS modeling to network optimization and multiple access \cite{dawy2017toward}.

In particular, one of the main challenges of cellular-enabled MTC in the IoT is the inability of existing random access (RA) protocols to support massive, short-packet transmissions. Moreover, the dense nature of MTCs will inevitably strain the highly-constrained resources of the RA process and, thus, render it inefficient. The RA challenges of MTC will be further exacerbated by the massive nature of the IoT which is expected to encompass around $3000$ MTDs within a geographically constrained area~\cite{dawy2017toward}, \cite{laya2014random}. Recently, there has been a surge in literature that focuses on optimizing RA process for MTC (e.g., see \cite{laya2014random} and references there in). Such works are primarily focused on either reducing signaling overhead to increase efficiency, or developing new backoff mechanisms to reduce collisions. However, solutions that focus optimizing the signaling overhead fall short in addressing the problem of resource congestion. Moreover, prior art \cite{laya2014random} that addresses the efforts to solve the RA channel congestion problem typically does so at the cost of increased latency. Such added latency cannot be sustained by mission-critical IoT applications that require reliable packet delivery within stringent deadlines. As a result, without discounting the existing efforts on improving RA for MTC, most of this prior art is still unsuitable to handle massive access due to the associated signaling overhead, collisions and delays.

Another promising approach to integrate the IoT into cellular systems is to use uncoordinated transmissions in which no RA procedure is performed and the MTDs are not scheduled \cite{UncoordinatedIoTMagazine}. In essence, for \emph{uncoordinated access}, MTDs select a random radio resource block (RB) and transmit their data. Even though this method reduces signaling, it still suffers from collisions since many MTDs might select the same RB. Despite some recent promising solutions for this uncoordinated access problem (e.g., see \cite{UncoordinatedIoTMagazine}), these existing approaches will still yield high congestion and associated delays.

Clearly, there is a need for new solutions for MTD access that can strike a balance between fully scheduled solutions, (that are controlled and reliable but have high signaling overhead, RA congestion and long delays) and fully uncoordinated solutions (that have low signaling overhead but have collisions, long delays, and increased receiver complexity).

The main contribution of this paper is therefore to develop such a middle-ground multiple access solution by leveraging the idea of a \emph{fast uplink grant}. The fast uplink grant is a method that was recently proposed by 3GPP  \cite{3GPP-fastuplinkgrant}, \cite{LTE14Outlook}. In the fast uplink grant scheme MTDs do not send RA scheduling requests and, instead, the BS will actively allocate uplink resources to those MTDs. Therefore, by using the fast uplink grant, MTDs are not required to perform RA, and, the problems associated with RA can be overcome. Moreover, in contrast to uncoordinated transmission, MTDs are scheduled by the BS and hence, collisions can be avoided.  To better understand the potential of this approach for the IoT, first, we provide an insight on on the opportunities provided by the use of the fast uplink channel for MTC. Then, we present an overview of the associated challenges, such as predicting which MTDs have data to transmit and properly scheduling those MTDs. To address these problems, we first exploit the potential of different learning methods for source traffic prediction. In this regard, we discuss a variety of tools and machine learning algorithms that can potentially be used to predict both periodic and event-driven MTC traffic. We then shed light on the use of multi-armed bandit (MAB) theory and deep reinforcement learning (Deep RL) as effective tools for enabling effective fast uplink grant allocation for massive MTC scenarios. To the best of our knowledge, this is the first work that analyzes how the fast uplink grant can be effectively leveraged to solve the emerging problem of massive MTCs in the IoT.
The rest of the paper is organized as follows. Section \ref{randomaccesssection} overviews the cellular RA process and its challenges for MTC. The fast uplink grant and its opportunities and problems are presented in Section \ref{fugtext}. A two-stage fast uplink grant approach for MTC is presented in Section \ref{twostage} and conclusions are drawn in Section \ref{conclutions}.

\section{Random Access for MTC: Overview and Challenges}\label{randomaccesssection}
\subsection{Overview of the RA Process}\label{rareivew}
The RA procedure is the first step needed to establish an uplink connection between any cellular device and a BS \cite{laya2014random}. Moreover, the RA process has two different forms, \emph{contention-based} and \emph{contention-free}. In contention-based RA, cellular users compete for RA resources and there is a possibility of collisions. In contrast, in contention-free case, the BS allocates specific resources for cellular users to send RA requests. For MTC, we focus on contention-based RA procedure because contention free RA is mostly suitable for transmitting emergency messages or very critical data, and the BS cannot reserve RA resources for all MTDs. The RA process in LTE/LTE-A systems has the following format. Upon having data to transmit, each user selects one RA slot, randomly from a set of available RA slots to send scheduling request. The number of RBs that are available for RA is limited since RA slots are allocated in uplink channel. In the frequency domain, each RA slot is $1.08$ MHz, which is equal to $6$ LTE RBs. In LTE each RB is  frequency-time unit one with $180$ kHz bandwidth and $1$ ms duration. In the time domain, the time intervals for RA availability vary between every $1$ ms to every $20$ ms, depending on the system configuration. There are a total of $64$ configurations which are broadcasted periodically by the BS. Once RA slots are available, cellular users randomly select one of the available RA slots and send the scheduling request in that slot. If the RA process is successful, the BS sends a RA response to the cellular user. However, if more than one device select the same RA slot for sending scheduling request, a collision occurs. The BS will attempt to decode the scheduling request in case of collision and will send an RA response which is received by all the users that had used the same RA slot. However, only the MTD whose data is successfully decoded at the BS is able to continue RA process and others are barred and have to send scheduling request in next RA opportunity. Once RA response is received, the cellular user will then transmit a connection request to the BS. In next step, the BS transmits contention resolutions and after that, cellular user starts transmitting data. Fig.~\ref{fug} illustrates a typical cellular RA process.

\subsection{Challenges of RA in MTC}\label{rachallengessubsection}
While the RA process of Section \ref{rareivew} is suitable for conventional human type devices, adopting it for MTC will face several challenges. For instance, the first challenge pertains to the limited number of RA opportunities in a cellular network. In fact, cellular RA resources are often much smaller than the anticipated number of MTDs in the IoT. For instance,  RA efficiency is maximized when the number of RA opportunities is equal to number devices competing for these resources. Increasing the number of RA slots is not feasible because RA slots are allocated in the physical uplink channel which has limited resources. Moreover, these resources are shared for both RA and uplink transmissions and, hence, there should be a balance between number of RBs allocated for the RA process and the number of resources left for uplink transmission. Hence, a small number of RA slots in comparison to the number of contending devices increases the probability of collisions. These collisions will make it impossible for the BS to decode RA pilots which will lead to a waste of resources and long delays for MTDs. The reason for long delays is that, after each RA failure, the affected MTD has to wait until the next RA opportunity to send the scheduling request again.

The second key RA challenge pertains to the short data packets size in MTC compared to conventional cellular services. For example, using six RBs for sending a scheduling request to transmit a short data packet that might require only one RB, is highly inefficient. The signaling overhead for RA is no longer negligible compared to the actual size of the data packets that will be transmitted by MTDs. Therefore, a low signaling scheduling scheme for MTC is highly desirable.

\subsection{Overview on Existing RA Solutions for MTC}
To address the aforementioned challenges of RA for MTC, several recent solutions have been proposed, as extensively reviewed in \cite{laya2014random}. The first class of solutions focus on coordinated transmissions. One popular solution is access class barring (ACB) in which the BS selects a number between zero and one and broadcasts it to the MTDs. Each MTD also randomly selects a number between zero and one. If the number that is selected by an MTD is smaller that the number that is sent by the BS, the MTD proceeds with RA. Otherwise, this MTD is barred from RA and waits until the next RA opportunity \cite{cooperativeACB}. While ACB solves the problem of RA channel congestion, it can potentially produce excessive delays due to the long waiting times experienced by barred MTDs. Another challenge in ACB is optimizing the value of the number that will be broadcast by the BS. To improve ACB, the work in \cite{extendedACB} introduced the notion of extended access class barring (EAB). The main premise of EAB is to bar  low-priority MTDs by pre-assigning different classes for MTDs and, hence, effectively improving system performance, in terms of heterogeneous QoS requirements. However, ACB schemes suffer from high RA signaling overhead. Another approach to improve RA for MTC, is to use the so-called access backoff process.  In this method, the BS encourages MTDs to not send a scheduling request for a time duration. However, by doing so, it increases latency. Another alternative solution for RA congestion is the notion of slotted RA in which each MTD is allocated a fixed RA opportunity to transmit only on that slot. However, slotted RA will not be suitable for massive MTC access since the periodicity of the RA slots will be large and, hence, incurring long delays. Moreover, if an MTD does not to send an RA pilot, the RA slot is wasted. Other methods have also been proposed for improving RA such as such as pull-based RA in which MTDs wait for permissions from BS to send RA pilots, priority-based RA in which specific RA priorities are assigned for devices. Code-expanded RA, self-optimizing load control, dynamic RA slot allocation and spatial grouping of MTDs. A comprehensive list of such methods and their advantages and disadvantages is presented in \cite{laya2014random}. However, all of these methods still exhibit high signaling overhead, collisions, and long delays which limits their applicability to MTC.

The second approach to solve the RA problem for MTC is to simply eliminate the RA process and use \emph{uncoordinated transmissions} \cite{UncoordinatedIoTMagazine}. Here, MTDs do not perform the RA process and, hence, they are not scheduled. Instead, the MTDs will select a random RB in uplink channel to transmit data \cite{UncoordinatedIoTMagazine}. Obviously, this method reduces the signaling level and can potentially increase the efficiency of the system. However, uncoordinated MTCs will still heavily suffer from collisions since the largely constrained uplink resources are shared among a potentially large number of MTDs that actively compete for channel access. Moreover, to realize uncoordinated MTC in practice, there is a need to design complicated receiver structures and retransmission mechanisms. Another major issue is scalability in terms of the number of supported devices. Since resources are selected randomly, efficiency is maximized when the number of devices is equal to number of resources. Clearly, in massive MTC, such requirement is not met, and, hence, the performance of the system will suffer. Code division multiple access (CDMA) is also investigated in \cite{cdmascalabilityforM2m} for MTC due to its support for heterogenous QoS requirements and scalability to support various number of devices. However, CDMA performance also suffers in the regime of large number of devices \cite{cdmascalabilityforM2m}. These drawback of uncoordinated transmissions limit the scope of their applicability to MTC, in general, and massive MTC, in particular.
\begin{table*}
\centering
\caption{Coordinated vs uncoordinated vs fast uplink grant}
\label{123table}
\begin{tabular}{|l|l|l|l|}
	\hline
	& \textbf{Signaling}                                                                     & \textbf{Collisions}                                                                                          & \textbf{Latency}                                                                                           \\ \hline
	\textbf{Coordinated}       & \begin{tabular}[c]{@{}l@{}}High (6 RBs) + \\ messages 2 to 4\end{tabular}              & \begin{tabular}[c]{@{}l@{}}High (number of MTDs $>>$ \\ number of RA slots)\end{tabular} & \begin{tabular}[c]{@{}l@{}}Waiting for RA slot + \\ RA signaling\end{tabular}                              \\ \hline
	\textbf{Uncoordinated}     & Zero                                                                                   & \begin{tabular}[c]{@{}l@{}}High(number of MTDs $>>$ \\ number of RBs)\end{tabular}       & \begin{tabular}[c]{@{}l@{}}High (when number of \\ MTDs $>>$ number of RBs)\end{tabular} \\ \hline
	\textbf{Fast uplink grant} & \begin{tabular}[c]{@{}l@{}}Small (one broadcast\\ message for entire cell)\end{tabular} & Zero                                                                                                         & \begin{tabular}[c]{@{}l@{}}Small (a few ms if grant\\ is allocated on time)\end{tabular}                   \\ \hline
\end{tabular}
\end{table*}

\section{Fast Uplink Grant}\label{fugtext}
\begin{figure}[t!]
  \centering
  \includegraphics[width=8cm]{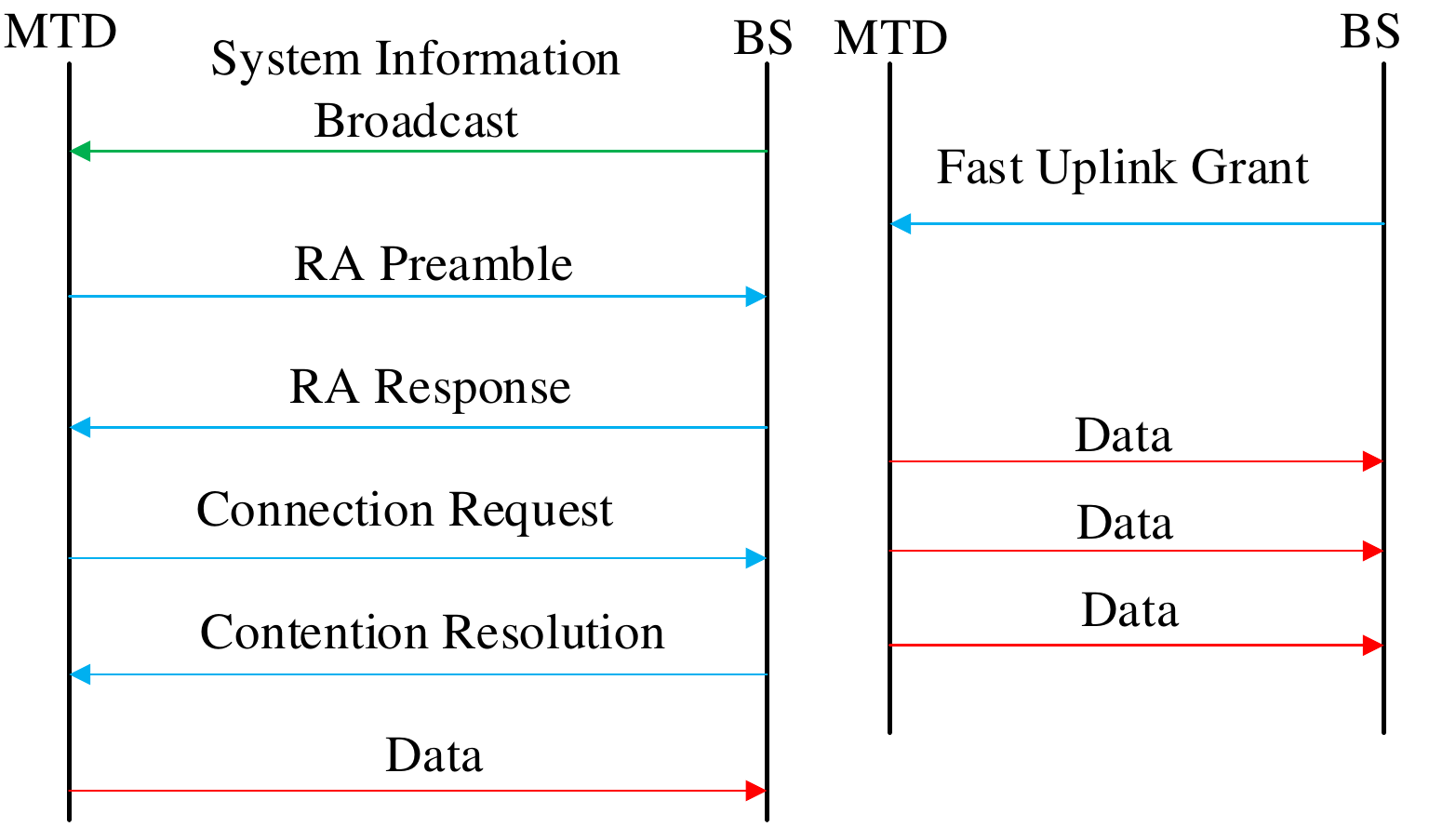}
  \caption{Comparison of the scheduling process in conventional RA (left) and in a fast uplink grant process (right).}\label{fug}
\end{figure}

A balanced approach between coordinated with RA process and uncoordinated transmissions could be developed by using the concept of a \emph{fast uplink grant}, as shown in Fig.~\ref{fug}.

\subsection{Fast Uplink Grant: Definition and Opportunities}
The fast uplink grant was introduced in \cite{3GPP-fastuplinkgrant}, \cite{LTE14Outlook} as an effective process that a cellular BS can use to select an MTD and allocate uplink resources to it. As such, by using the fast uplink grant, the MTDs will no longer need to perform a RA process. Instead, whenever an MTD has data transmit, it can simply wait for the fast uplink grant to be allocated to it.

The fast uplink grant presents several benefits compared to the schemes presented in previous sections. First, the amount of signaling that is required is much less than RA. This is due to the fact that, by using the fast uplink grant:
\begin{itemize}
  \item Only one level of signaling is performed.
  \item The amount of signaling is smaller since the fast uplink grant for the entire system can be sent in one broadcast message.
\end{itemize}
Second, in a system with large number of devices, collisions of RA pilots in coordinated access and packet collisions in uncoordinated transmission can be overcome by using the fast uplink grant. Also, implementing the fast uplink grant will not require complicated receiver structures at the BS to solve problems of uncontrolled transmission. The benefits of fast uplink grant can be summarized as follows:
\begin{itemize}
  \item RA congestion is mitigated.
  \item RA radio resources can be used to transmit uplink data and hence, larger number of devices can be supported.
  \item Packet collisions of uncoordinated transmission are avoided.
  \item The BS can satisfy the heterogenous QoS requirements of MTDs by active allocation of fast uplink grant to MTDs with stricter latency requirements.
   \item MTDs can save energy by skipping RA process and retransmission of scheduling requests in case of RA failure.
   \item RA prcess delay and delay of waiting for next RA opportunity in case of collisons are eliminated.
\end{itemize}
A summary of differences between fast uplink grant with conventional schems is given Table \ref{123table}.

\subsection{Challenges of Fast Uplink Grant}\label{fugchallenges}
The first drawback of the fast uplink grant is the possibility of wasting resources whenever an MTD that has received fast uplink grant does not have data to transmit \cite{LTE14Outlook}. Moreover, the if fast uplink grant is not received within the maximum tolerable delay of the data packets, then packet will be dropped yielding transmission failures. This can potentially lead to major problems for both latency and reliability in MTC. In essence, to adopt the fast uplink grant for MTC, one of the main challenges that must be overcome is the optimal selection of MTDs by the BS. This MTD selection process, in turn, faces two key challenges. First, there is a need to predict the set of nodes that will likely have data to transmit, at any given time. By doing accurate predictions, the BS can solve the problem of allocating fast uplink grant to silent MTDs. Once predictions are properly implemented, the BS must also be able to determine the scheduling sequence of MTDs. This challenge is particularly pronounced when the number of devices significantly exceeds the number of resources. Hence, sophisticated scheduling algorithms are required to be developed to enable fast uplink grant allocation. In what follows, we propose a two-stage approach for leveraging the fast uplink grant for MTCs.

\section{Proposed Two-Stage Solution}\label{twostage}
\subsection{Source Traffic Prediction in MTC}\label{traficprediction}
As stated in previous section, if an MTD is selected for fast uplink grant and does not have data to transmit, uplink radio resources are wasted. To address this challenge, the BS must implement advanced traffic prediction mechanisms to predict the set of MTDs that have data to transmit. Most of the prior art on traffic modeling for MTC is focused on aggregate traffic modeling at the BS. Such traffic modeling only estimates the number of devices or number of packets arriving in the system. However, source traffic modeling is fundamentally a different problem since we are interested in precisely predicting which MTDs will enter the network. In essence, at each time slot, the BS needs to predict which MTDs will have traffic to send and, hence, need uplink resources. Such predictions are generally feasible in an IoT environment due to two facts: a) most of the MTDs are stationary or exhibit low mobility and b) The set of MTDs communicating with a BS is often fixed.

For traffic predictions, one needs to distinguish between two types of MTC traffic: periodic reporting and event-driven transmissions. In periodic reporting, MTDs periodically transmit data packets at specific, pre-determined times. In event-driven traffic, often, a large number of MTDs will initiate a transmission request to provide reports on a certain IoT event. Clearly, prediction of event-driven traffic is much harder than periodic traffic. Next, we present mathematical tools that can be used to develop algorithms for prediction of both MTC traffic types.

\subsubsection{Prediction of Periodic Traffic}
Many IoT applications, such as smart metering and weather sensing, rely on MTDs that \emph{periodically} transmit sensory data generated from the observations of the physical environment. Different applications generate heterogeneous data sizes in various period durations. These durations could be as low as a few milliseconds and up to once in a month. An MTD might also transmit data pertaining to multiple IoT appliations. This results in differenet data packets with different transmission intervals. Hence, the BS must learn exact time instances at which any given MTD will generate its data, as well as the associated packet size. Clearly, the BS must collect data from the past transmissions of all MTDs and subsequently use machine learning algorithms to predict the source traffic for each MTD. This prediction must be precise, since some IoT applications generate data with very strict latency requirements, as low as $10$ ms. Mathematical methods such as non-homogeneous Poisson Process (NHPP) could be used to model the arrival rate of packets to the queue of each MTD at different times. In NHPP, arrivals follow Poisson distribution, however, at each time, the rate of arrival is different. Such pattern analysis is called calendar-based periodic pattern mining and models such as the sequential association rule and the calendar association rule exist for analyzing them (e.g, see\cite{periodicpattern} and reference therein).

\subsubsection{Prediction of Event-Driven Traffic}
In IoT applications that rely on event-driven MTC, whenever an event occurs, several MTDs that detect the event must initiate data transmission to the BS. This leads to a burst of RA scheduling requests from a large number of MTDs. Such event-driven MTC traffic will exacerbate the challenges pertaining to scheduling a large number of MTDs that were identified in Section \ref{rachallengessubsection}. Hence, effective traffic prediction in event-driven MTC is critical. Naturally, predicting an IoT event that was never observed is not possible. However, it is possible to detect an event based on unusual traffic generated by MTDs. If the BS, based on the data gathered from previous IoT events, can calculate the likelihood with which other MTDs face the same event, it will be possible to design algorithms to predict event-driven traffic. For example, a predictive grant allocation is presented in \cite{brownPredictive} for IoT cases in which an event propagates through a system of sensors located in a line. Here, the BS can learn which MTDs will initiate transmission in case of an event. However, this method cannot be generalized to all possible types of MTC events.

\begin{figure*}[ht]
	\centering
	\includegraphics[width=13cm]{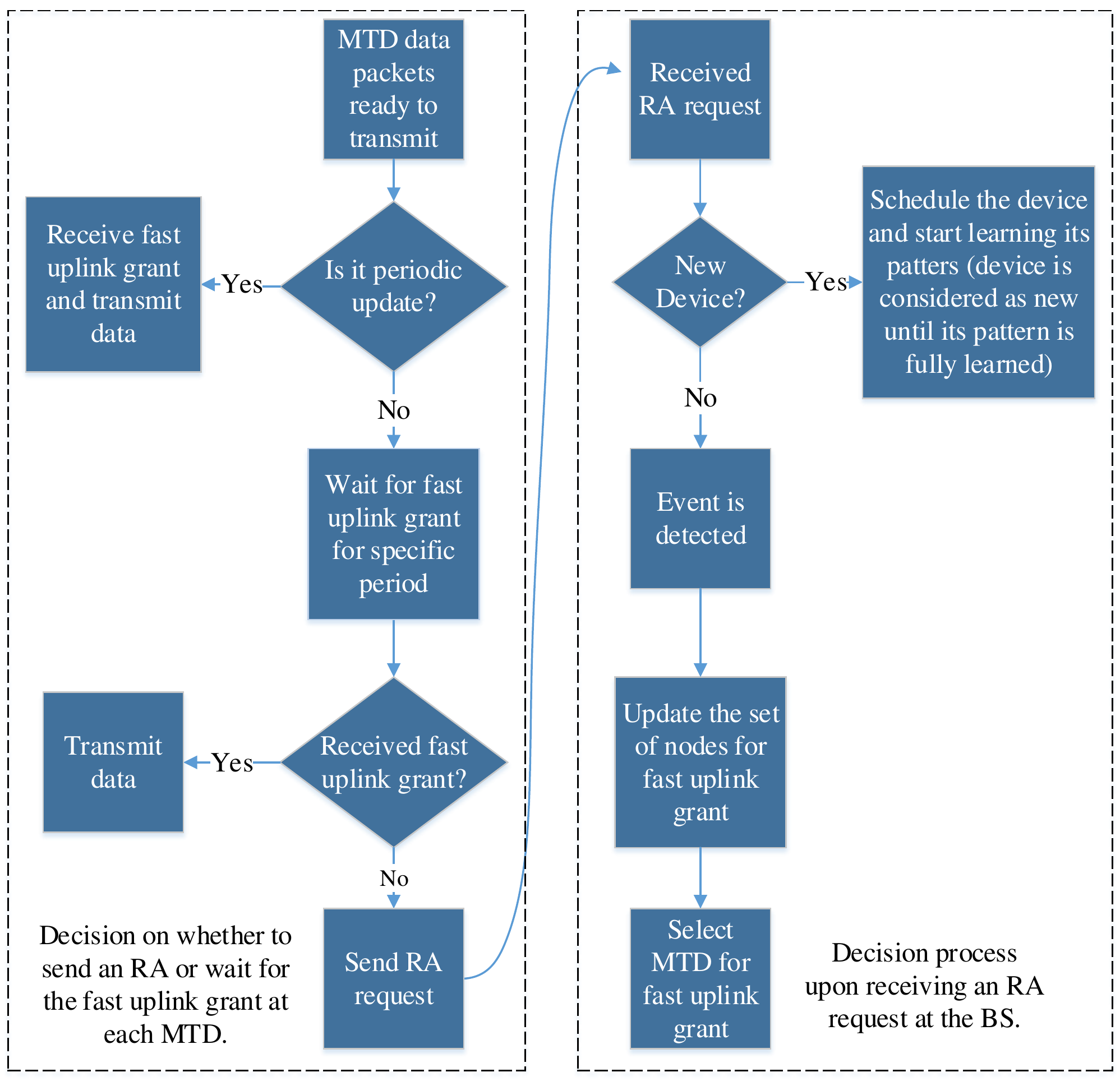}
	\caption{Flowchart of the proposed algorithm for RA decision making as implemented by any given MTD (left) and event detection and uplink grant allocation at the BS (right).}\label{flowchart}
\end{figure*}

Here, we present novel methods that could be used for source traffic prediction in event-driven MTCs. First, we assume that, during past events, the BS has collected the data about the transmission of MTDs. That is, the BS knows which devices were transmitting during each event along with their order of transmission. Second, we assume that the set of MTDs with periodic traffic is predicted and MTDs do not send scheduling requests for periodic reporting. Hence, any scheduling request can be considered as event trigger and used for detection of events. We could also consider that, once an event happens, MTDs wait for a short period of time for an uplink grant, if they do not receive it, they use RA. A flowchart of decision making at MTD for RA is given in Fig. \ref{flowchart}. Now, once an event happens, some MTDs will report it earlier than others. The BS considers the first RA request as event trigger. The event-driven traffic prediction problem is now simplified to the following question: \emph{Once a specific MTD detects an event, which other MTDs will experience the same event with a high probability?} Answering this question requires analysis of the data collected from previous events. One natural solution here is to use probabilistic models from machine learning. Using the previously collected data, probabilistic relationship between two MTDs facing the same event can be calculated. Another possible solution could be to use the paradigm of \emph{causality}. Causality deals with the following problem: Given than an MTD detects an event, which other MTDs that specific MTD \emph{statistically causes} to detect the same event. Granger causality \cite{granger1969investigating} is one method that is used in machine learning to investigate the causality between two sequences of random variables. Another advanced and novel method is based on \emph{directed information} \cite{massey1990causality}. Directed information can be used to infer causality between sequences of random variables. Considering two sequences of random variables, past and present values of the first sequence, and past values of the second sequence can be used to evaluate the present value of second sequence. Directed information is a powerful method that is used for prediction of seizure in Epliepsy patients and causality between neurons of the human brain. Hence, such a framework could potentially be used to predict the exact number of packets that an MTD has to transmit in case of an event. Once causality is inferred, one can predict which MTDs face the same IoT event and start allocating fast uplink grant to them. In Fig. \ref{flowchart}, we present the flowchart of an algorithm to that can use event-detection in the BS for fast uplink grant allocation.
\subsection{Optimal Fast Uplink Grant Allocation}\label{optimizing}
Once the set of MTDs that have data to transmit is predicted, the network must select which ones can be granted access. Due to the limited radio resources, MTDs should be scheduled based on their QoS requirements, particularly the maximum tolerable delay of their packets. If fast uplink grants are allocated randomly, it is possible that the network may prioritize the scheduling of delay-tolerant MTD data, thus jeopardizing the performance of delay-sensitive MTD data. If the BS has full knowledge of the QoS requirements of all MTDs, this scheduling can be performed in a centralized manner. Such scheduling could take into account performance indicators like latency, value of data packets, and the wireless channel quality. However, in a realistic scenario, such information might not be available to the BS and any fast uplink grant allocation algorithm should be able to select MTDs in an uncertain environment. Therefore, the design of sophisticated algorithms for optimal fast uplink grant allocation is needed. Here, we present some initial directions toward building such scheduling algorithms that exploit recent advances in machine learning and artificial intelligence to optimize the allocation of fast uplink grants to MTDs \cite{walidML}.

\subsubsection{Multi-Armed Bandit Theory}
Multi-armed bandits (MABs) are a class of reinforcement learning (RL) problems that deal with decision making in uncertain environments with limited or no prior information \cite{sutton1998reinforcement}. The basic MAB problem consists of a set of arms (available actions) that can be chosen by a decision-making agent that plays an arm at each time and receives a reward. The rewards are drawn from an unknown probability distribution. The agent has no prior information about the rewards of each arm and has to randomly select arms, observe the rewards, and, then, try to find the best possible arm. In MAB, the notion of \emph{regret} -- defined as the difference between the best possible arm that could have been played and the arm that is selected -- is used as a measure of performance. The main goal of any MAB algorithm is to minimize the cumulative regret over time. To solve conventional MAB problems, algorithms such as $\epsilon$-greedy and upper confidence bound (UCB) are often adopted. There are also special MAB problems such as \emph{sleeping MAB} problems in which, at each time, only a subset of arms are available for the agent, or \emph{contextual MAB} where at each time there is some side information provided to the decision maker. Clearly, such problems are apropos for addressing the MTD selection problem. For example, the sleeping MAB framework is particularly suitable to select MTDs for fast uplink grant, since the availability of MTDs can change at each time. Moreover, rewards in MAB setting can be defined in terms of various QoS metrics, and, hence, the algorithms can be used to optimize the selected metric.

\subsubsection{Deep Reinforcement Learning}
Deep RL is used in RL problems with extremely large states and actions where it is not possible to explore all the possible states and actions. Clearly, MTC scheduling problems with a large number of MTDs will have to deal with such large action and state spaces. In deep RL, neural networks (NN) are used to approximate the environment, and, for the states that were not seen before, the NN output determines the action \cite{walidML}. To use deep RL for MTD selection using the fast uplink grant, one can first formulate the problem using a Markov decision process (MDP) \cite{sutton1998reinforcement}. In this MDP formulation, each state is a combination of set of available MTDs and their associated QoS requirements and each action of MDP is a subset of set of available MTDs. Each action will move the system to a new state, that is the new set of available MTDs. To find the optimal action for each given set, one can use Deep RL algorithms. The states and actions are explored to train the NN and for new states, a Deep RL algorithm can find the optimal action. The key advantage of Deep RL over other possible scheduling methods is that after training, it requires low computation, and, it is suitable for online decision making problems.

\section{Conclusion}\label{conclutions}
In this paper, we have studied the potential of using the fast uplink grant as an enabler for massive MTCs in the IoT. First, we have reviewed the challenges that conventional access schemes face in MTC, and, discussed the fast uplink grant as a solution. Then, we have presented challenges of the fast uplink grant, and, proposed solutions to address them. In particular, we have presented methods for source traffic prediction, for both periodic and event-driven traffic. Then, we have proposed machine learning methods for optimal selection of MTDs for fast uplink grant. In a nutshell, this work provides a stepping stone towards a better understanding of how the notion of a fast uplink grant can be effectively leveraged for massive MTCs.

\bibliographystyle{IEEEtran}
\bibliography{diIntDiv}
\end{document}